# Experimental study and critical review of structural, thermodynamic and mechanical properties of superhard refractory boron suboxide $B_6O$

**Oleksandr O. Kurakevych** [a] and **Vladimir L. Solozhenko** [b]

[a] CMCP & IMPMC, Université P & M Curie, 75015 Paris, France
[b] LSPM–CNRS, Université Paris Nord, 93430 Villetaneuse, France

*In the present paper we performed the analysis of available data on structural, thermodynamic and mechanical properties of $B_6O$. Although the compound is known for half a century and has been extensively studied, many properties of this boron-rich solid remain unknown or doubtful. Semi-empirical analysis of our experimental and literature data allowed us to choose the best values of main thermodynamic and mechanical characteristics among previously reported data, to predict the thermoelastic equation of state of $B_6O$, and dependence of its hardness on non-stoichiometry and temperature.*

**Keywords:** superhard materials, boron suboxide, thermodynamics, crystal structure, mechanical properties.

Boron and boron-rich compounds are strategic materials for nuclear, superabrasive and advanced electronic applications [1-4]. Boron suboxide, $B_6O$ is a boron-rich solid known already for half a century [5]. It attracted much attention due to its mechanical [6], structural [7], phononic [8], etc., properties. However, many properties of $B_6O$ remain poorly understood to present time, similarly to other boron-rich solids with α-$B_{12}$ structural type (such as pristine α-$B_{12}$ [9], $B_{4+x}C_{1-x}$ [10], $B_{13}N_2$ [11,12] and $B_{12}X_2$ phases where X – P, As [8]).

**Synthesis**

At ambient pressure $B_6O$ can be synthesized by the oxidation of boron with boron oxide $B_2O_3$ [5] or some metal oxides such as ZnO, MgO, SnO, CdO [13,14] at 1500-1900 K in argon atmosphere. However, the samples prepared at ambient pressure suffer from remarkable non-stoichiometry [15].

The procedure of high-pressure synthesis of $B_6O$ by interaction of amorphous boron with $B_2O_3$ in a multianvil apparatus was developed by Hubert et al. [16,17]. It has been found that ideal conditions for the synthesis of $B_6O$ crystals of deep red color and icosahedral habit [7] are 4-5.5 GPa and 2000-2100 K. The best stoichiometry ($B_6O_x$ with $x$ up to 0.95) has been achieved under pressures of ~6 GPa. Later it has been shown that the synthesis of the almost stoichiometric phase is possible already at pressures as low as 1 GPa, when crystalline β-$B_{106}$ is used as starting material instead of amorphous boron [18].

**Structural and lattice-vibration properties**

Boron suboxide B$_6$O, similarly to α-B$_{12}$ and B$_4$C, crystallizes in the trigonal syngony, *R*-3*m* space group [19,20]. The values of lattice parameters extrapolated to ideal stoichiometry are *a* = 5.399 Å, *c* = 12.306 Å for hexagonal setting. Lattice parameters are very sensitive to the O-vacancies (1–*x*) in the ideal B$_6$O lattice (Fig. 1a,b). Assuming the B$_6$O$_x$ composition, *da*/*dx* = 0.083 Å and *dc*/*dx* = -0.117 Å [16].

The Raman spectrum of boron suboxide is rather complicated. When conventional green or red lasers are used, it cannot be observed due to the strong fluorescence [8,18]. Blue, IR or UV laser excitation beams should be used to obtain the Raman spectra of B$_6$O (Fig. 1c) [18]. The first order Raman spectrum contains 11 well-resolved lines of the 12 expected modes 5 A$_{1g}$ + 7 E$_g$ for space group R-3m. The second order Raman spectrum contains 8 bands that are resolved only in the case of the 244-nm excitation line [18]. Although the attribution of Raman modes has not been unambiguously made so far because of the absence of single-crystal data, the principal Raman features were studied in details [8,18]. All the lines below 600 cm$^{-1}$ have been attributed to vibrations with participation of oxygen atoms, while all the lines above, to vibrations involving only boron atoms in icosahedra (intra- and intericosahedral vibrations). The fine feature just below 500 cm$^{-1}$ corresponds to the symmetric stretching of the O–O pairs; and the narrow line just above 500 cm$^{-1}$, to a motion of icosahedral boron atoms about oxygen atoms.

**Melting and thermal stability**

At ambient pressure B$_6$O decomposes above 2030 K. The products of decomposition have not been unambiguously identified, and, most probably, contain boron and O$_2$ or gaseous BO [5]. At the same time, at high pressures B$_6$O melts congruently [21]. At 5.8 GPa, the melting of B$_6$O phase was observed at 2710±40 K, which was accompanied by the appearance of a characteristic diffuse halo in the X-ray powder diffraction pattern. The cooling of the melt down to 2100 K at a rate of ~10 K/s was accompanied by the crystallization of B$_6$O, while the fast quenching leads to the recovery of crystalline boron (β-B$_{106}$) and B$_2$O$_3$. At 4.3 GPa, B$_6$O melts at temperature of about 2620 K, thus, the melting curve has a positive slope of ~60 K/GPa [21]. The melting curve of B$_6$O is presented in Fig. 2a.

The phase equilibria with participation of B$_6$O in the B–B$_2$O$_3$ system at 5 GPa has been determined based on *in situ* experiments in the recent study by Solozhenko et al. [22], while fitting the experimental *p-T-x* data to phenomenological equations describing free enthalpy of solids and liquid phase allowed to construct the phase diagram of this system (see Fig. 2b). At high pressure B$_6$O melts congruently, and forms two eutectic equilibria with α-B$_2$O$_3$ and β-B$_{106}$, respectively. The boron-rich part of the phase diagram (especially, below melting) has not been unambiguously studied so far.

**Thermodynamic properties.**

The standard enthalpy of formation of $B_6O$ can be estimated as $\Delta_f H° = -1244.5$ kJ/mol [17]. To present time, only two sets of experimental data on the heat capacity of $B_6O$ are known [23], i.e. one at low temperatures down to 10 K, and another at high temperatures up to 800 K. Here, we have fitted both sets to Holzapfel's adaptive pseudo-Debye equation [24], i.e.

$$C_V = 3R\tau^3 \frac{4C_0 + 3C_1\tau + 2C_2\tau^2 + C_3\tau^3}{(C_0 + C_1\tau + C_2\tau^2 + C_3\tau^3)^2}\left[1 + A\frac{\tau^4}{(a+\tau)^3}\right], \qquad (1)$$

where $\tau = T/\theta_h$; $\theta_h$ is Debye temperature in the high-temperature region; $C_1$, $C_2$ and $A$ are parameters to be fitted; $C_3 = 1$; $a$ characterizes non-harmonicity and was fixed to 1/8; R is gas constant. Parameter $C_0$ has been chosen as $C_0 = (5\,\theta_l^3)/(\pi^4\,\theta_h^3)$, with $\theta_l$ and $\theta_h$ Debye temperatures in the low- and high-temperature regions, respectively, in order to obtain these values directly as parameters of fitting. The results of the data fit (Tab. 1, Fig. 3a) give the $\theta$ values of the same order of magnitude as for other boron-rich solids with similar structure ($B_4C$, $\alpha$–$B_{12}$). The corresponding normalized standard values of heat capacity, enthalpy and entropy are also very close. The temperature dependencies of enthalpy and entropy of $B_6O$ are presented in Fig. 3b.

In order to simulate the thermoelastic data, we have used the Anderson-Grüneisen model [25], i.e.

$$\alpha(p,T) = \alpha(0,T)\left(\frac{V(p,T)}{V(0,T)}\right)^{\delta_T} \qquad (2)$$

or, in the terms of bulk modus,

$$B(p,T) = B(p,300)\left(\frac{V(p,T)}{V(p,300)}\right)^{-\delta_T} \qquad (3)$$

Under assumption that $\delta_T$ is constant and taking into account the definition of thermal expansion, (2) can be easily transformed into

$$\int_{T_1}^{T_2} d[V(p,T)]^{-\delta_T} = \int_{T_1}^{T_2} d[V(0,T)]^{-\delta_T}, \qquad (4)$$

and, therefore,

$$V(p,T) = \left[V(0,T)^{-\delta_T} + V(p,300)^{-\delta_T} - V(0,300)^{-\delta_T}\right]^{-1/\delta_T}. \qquad (5)$$

During fitting procedure, the $V(0,T)$ dependence was suggested to follow equation

$$V(0,T) = V(0,300\text{ K})\,[1 + a\,(T-300) + b\,(T-300)^2], \qquad (6)$$

while the $V(p, 300\,\text{K})$ dependence has been fixed to the 300-K equation of state of $B_6O$ reported in [26]. In the first approximation [27-29] one may suggest that $\delta_T \approx B' = 6$ [26]. Fitting of all available experimental $p$-$V$-$T$ data (at 0.1 MPa according to Refs. 30,31, at 5.8 GPa using the experimental points obtained during the study of melting [21], Fig. 3c) to Eq. (5) leads to $a = 1.5 \cdot 10^{-5}\,\text{K}^{-1}$, $b = 5 \cdot 10^{-9}\,\text{K}^{-2}$, while $V(p, 300\,\text{K})$ is defined by Vinet equation of state with $B_0 = 180$ GPa and $B_0' = 6$ [26] ($V_0 = 314.5\,\text{Å}^3$).

**Mechanical properties.**

$B_6O$ is a superhard phase, much harder then other boron oxides [32]; and, probably, is the hardest known oxide [33]. Anyway, the hardness of $B_6O$ still remains the subject of discussion. The reported values of Vickers hardness $H_V$ for polycrystalline $B_6O$ ceramics may vary from 30 to 45 GPa [6]. Tab. 2 summarizes mechanical properties of $B_6O$ in comparison with $B_4C$ and hard high-pressure phase of $B_2O_3$ ($\beta$-$B_2O_3$).

When using the thermodynamic model of hardness [28,29,34,35], the calculated value of $H_V$ for $B_6O$, i.e. 37.3 GPa, is in a very good agreement with the experimental value of $H_V = 38$ GPa [5] for well-sintered polycrystalline $B_6O$ with density close to the crystallographic one. This is the only experimental value obtained in the load-independent $H_V$ region (applied force of 10 N) recommended for estimation of hardness of superhard materials [36]. The lower value of hardness for $B_6O$ as compared to $B_4C$ ($H_V = 45$ GPa for single crystal [37]) may be explained by the higher ionicity of the B-O bonds as compared to B-C bonds [35]. Thus, the value of $H_V = 38$ GPa seems to be the most reasonable estimate for the $B_6O$ hardness, and well agrees with *ab initio* simulations of stress-strain curves for $B_6O$ crystal faces (see, for example, paper by Veprek et al. [38]).

For simulation of $B_6O_x$ hardness as a function of composition and temperature, we have used equation [28,35]

$$H_V = \frac{2\Delta G^\circ_{at}}{VN}\alpha\beta\varepsilon, \quad (7)$$

where $V = V(x,T)$ is molar (atomic) volume (cm$^3$ mole$^{-1}$); $N$ is maximal coordination number[1]; $\alpha = \alpha(T)$ is coefficient of relative (as compared to diamond) plasticity; $\beta$ is coefficient corresponding to the bond polarity (see below); $\varepsilon$ is ratio between the mean number of valent electrons per atom and the number of bonds with neighboring atoms $(N)$[2]; $\Delta G^\circ_{at} = \Delta G^\circ_{at}(x,T)$ is standard Gibbs energy of atomization of compound (kJ mole$^{-1}$) (see [34] for the details).

One can clearly see that even in the case of strong non-stoichiometry (50% of O-places occupied in the lattice), the material is expected to lose only ~10% of its initial hardness (Fig. 4a). In the framework of this model [28,34] the bulk modulus should closely follow the similar dependence (Fig. 4b), according to equation

---

[1] For some compounds of very complex structure, such as boron-rich solids, we have used a mean/effective value.

[2] The use of this coefficient allows one to establish the hardness of the A$^I$B$^{VII}$ ($\varepsilon = 1/N$) and A$^{II}$B$^{VI}$ ($\varepsilon = 2/N$) compounds. In the case of $B_6O$, $\varepsilon = 1$.

$$H_V = \frac{2}{3}\frac{g\alpha\varepsilon\sqrt{\beta}}{N}B, \qquad (8)$$

where *g* is a constant for a given class of compounds [28].

The temperature has much higher impact on hardness (Fig. 4c), but even at ~1600 K the hardness of B$_6$O should still remain at the level of the WC–10%Co hard alloy (16 ГПа) [39].

**Conclusions.**

Boron suboxide B$_6$O with stoichiometry close to the ideal one can be synthesized already at about 1 GPa in the case of oxidation of crystalline boron. At high pressures the compound melts congruently, and its thermodynamic properties are similar to those of other boron-rich solids with the structure related to α-B$_{12}$ phase. Although the mechanical properties of boron suboxide have not been unambiguously studied so far, the most reasonable hardness value seems to be $H_V$ = 38 GPa.

**References.**

1.  *Oganov, A.R., Chen, J., Gatti, C., et al.* Ionic high-pressure form of elemental boron. // Nature - 2009. - **457**, N 7231. - P. 863-867.

2.  *Oganov, A.R., Chen, J., Gatti, C., et al.* Addendum: Ionic high-pressure form of elemental boron. // Nature - 2009. - **460**, N 7252. - P. 292-292.

3.  *Oganov, A.R., Solozhenko, V.L.* Boron: a hunt for superhard polymorphs. // J. Superhard Mater. - 2009. - **31**, N 5. - P. 285-291

4.  *Kurakevych, O.O.* Superhard phases of simple substances and binary compounds of the B-C-N-O system: from diamond to the latest results (a Review) // J. Superhard Mater. - 2009. - **31**, N 3. - P. 139-157.

5.  *Rizzo, H.F., Simmons, W.C., Bielstein, H.O.* The existence and formation of the solid B$_6$O. // J. Electrochem. Soc. - 1962. - **109**, N 11. - P. 1079-1082

6.  *He, D., Zhao, Y., Daemen, L., et al.* Boron suboxide: As hard as cubic boron nitride. // Appl. Phys. Lett. - 2002. - **81**, N 4. - P. 643-645.

7.  *Hubert, H., Devouard, B., Garvie, L.A.J., et al.* Icosahedral packing of B-12 icosahedra in boron suboxide (B$_6$O). // Nature - 1998. - **391**, N 6665. - P. 376-378.

8.  *Aselage, T.L., Tallant, D.R., Emin, D.* Isotope dependencies of Raman spectra of B$_{12}$As$_2$, B$_{12}$P$_2$, B$_{12}$O$_2$, and B$_{12+x}$C$_{3-x}$: Bonding of intericosahedral chains. // Phys. Rev. B - 1997. - **56**, N 6. - P. 3122-3129.

9.  *Decker, B.F., Kasper, J.S.* The crystal structure of a simple rhombohedral form of boron. // Acta Crystallogr. - 1959. - **12**, N 7. - P. 503-506.

10. *Thevenot, F.* Boron carbide - A comprehensive review. // J. Europ. Ceram. Soc. - 1990. - **6**, N 4. - P. 205-225.

11. *Kurakevych, O.O., Solozhenko, V.L.* Rhombohedral boron subnitride, B$_{13}$N$_2$, by X-ray powder diffraction. // Acta Crystallogr. C - 2007. - **63**, N 9. - P. i80-i82.

12. *Solozhenko, V.L., Kurakevych, O.O.* Chemical interaction in the B–BN system at high pressures and temperatures. Synthesis of novel boron subnitrides. // J. Solid State Chem. - 2009. - **182**, N 6. - P. 1359-1364.

13. *Holcombe, C.E., Horne, O.J.* Preparation of boron suboxide, $B_7O$. // J. Amer. Ceram. Soc. - 1972. - **55**, N 2. - P. 106-106.

14. Holcombe, J., C. E. , Horne Jr., O.J. (1971-1972) US Patent 3-660-031.

15. *Olofsson, M., Lundstrom, T.* Synthesis and structure of non-stoichiometric $B_6O$. // J. Alloy. Comp. - 1997. - **257**, N 1-2. - P. 91-95.

16. *Hubert, H., Garvie, L.A.J., Devouard, B., et al.* High-pressure, high-temperature synthesis and characterization of boron suboxide ($B_6O$). // Chem. Mater. - 1998. - **10**, N 6. - P. 1530-1537.

17. *McMillan, P.F., Hubert, H., Chizmeshya, A., et al.* Nucleation and growth of icosahedral boron suboxide clusters at high pressure. // J. Solid State Chem. - 1999. - **147**, N 1. - P. 281-290.

18. *Solozhenko, V.L., Kurakevych, O.O., Bouvier, P.* First and second order Raman scattering of $B_6O$. // J. Raman Spectr. - 2009. - **40**, N 8. - P. 1078-1081.

19. *Higashi, I., Kobayashi, M., Bernhard, J., et al.* Crystal structure of $B_6O$. // AIP Conference Proceedings - 1991. - **231**. - P. 201-204.

20. *Kobayashi, M., Higashi, I., Brodhag, C., Thevenot, F.* Structure of $B_6O$ boron suboxide by Rietveld refinement. // J. Mat. Sci. - 1993. - **28**. - P. 2129-2134.

21. *Solozhenko, V.L., Lathe, C.* On the melting temperature of $B_6O$. // J. Superhard Mater. - 2007. - **29**, N 4. - P. 259–260.

22. *Solozhenko, V.L., Kurakevych, O.O., Turkevich, V.Z., et al.* Phase diagram of the $B–B_2O_3$ system at 5 GPa: Experimental and theoretical studies. // J. Phys. Chem. B - 2008. - **112**, N 21. - P. 6683-6687.

23. *Tsagareishvili, G.V., Tsagareishvili, D.S., Tushishvili, M.C., et al.* Thermodynamic properties of boron suboxide in the temperature range 11.44-781.8 K. // AIP Conference Proceedings - 1991. – **231**. - P. 384-391.

24. *Holzapfel, W.B.* Equations of state for solids under strong compression. // High Pressure Res. - 1998. - **16**, N 2. - P. 81-126.

25. *Anderson, O.L., Isaak, D., Oda, H.* High-temperature elastic constant data on minerals relevant to geophysics. // Rev. Geophys. - 1992. - **30**, N 1. - P. 57-90.

26. *Nieto-Sanz, D., Loubeyre, P., Crichton, W., et al.* X-ray study of the synthesis of boron oxides at high pressure: phase diagram and equation of state. // Phys. Rev. B - 2004. - **70**, N 21. - P. 214108 1-6.

27. *Shirai, K., Masago, A., Katayama-Yoshida, H.* High-pressure properties and phase diagram of boron. // Phys. Stat. Solidi B - 2007. - **244**, N 1. - P. 303-308.

28. *Mukhanov, V.A., Kurakevych, O.O., Solozhenko, V.L.* The interrelation between hardness and compressibility of substances and their structure and thermodynamic properties. // J. Superhard Mater. - 2008. - **30**, N 6. - P. 368-378.

29. *Mukhanov, V.A., Kurakevych, O.O., Solozhenko, V.L.* Thermodynamic aspects of materials' hardness: prediction of novel superhard high-pressure phases. // High Press. Res. - 2008. - **28**, N 4. - P. 531-537.

30. *Tushishvili, M.C., Tsagareishvili, G.V., Tsagareishvili, D.S.* Thermoelastic properties of boron suboxide in the 0-1500 K range. . // J. Hard Mater. - 1992. – **3**. - P. 225-233.

31. *Tsagareishvili, D.S., Tushishvili, M.C., Tsagareishvili, G.V.* Estimation of some thermoelastic properties of boron suboxide in wide ranges of temperature. // AIP Conference Proceedings - 1991. - **231**, N 1. - P. 392-395.

32. *Mukhanov, V.A., Kurakevich, O.O., Solozhenko, V.L.* On the hardness of boron (III) oxide. // J. Superhard Mater. - 2008. - **32**, N 1. - P. 71-72.

33. *Oganov, A., Lyakhov, A.* Towards the theory of hardness of materials. // J. Superhard Mater. - 2010. - **32**, N 3. - P. 143-147.

34. *Mukhanov, V.A., Kurakevych, O.O., Solozhenko, V.L.* Hardness of materials at high temperature and high pressure. // Philosoph. Mag. - 2009. - **89**, N 25. - P. 2117 - 2127.

35. *Mukhanov, V.A., Kurakevych, O.O., Solozhenko, V.L.* Thermodynamic model of hardness: Particular case of boron-rich solids. // J. Superhard Mater. - 2010. - **32**, N 3. - P. 167-176.

36. *Brazhkin, V., Dubrovinskaia, N., Nicol, M., et al.* What does "harder then diamond" mean? // Nature Mater. - 2004. - **3**, N - P. 576-577.

37. *Domnich, V., Gogotsi, Y., Trenary, M.* Identification of pressure-induced phase transformations using nanoindentation. // Mater. Res. Soc. Symp. Proc. - 2001. – **649**. - P. Q8.9.1-Q8.9.6.

38. *Veprek, S., Zhang, R.F., Argon A.S.* Mechanical properties and hardness of boron and boron-rich solids. // J. Superhard Mater. - 2011. - ***this volume***, N *this issue*. - P. *xx-xx*.

39. *Schubert, W.D., Neumeister, H., Kinger, G., et al.* Hardness to toughness relationship of fine-grained WC-Co hardmetals. // Int. J. Refract. Met. Hard Mater. - 1998. - **16**, N 2. - P. 133-142.

40. *Tsagareishvili, D.S., Tsagareishvili, G.V., Omiadze, I.S., et al.* Thermodynamic properties of α-rhombohedral boron from 16.05 to 714.5 K. // J. Less Comm. Met. - 1986. - **117**, N 1-2. - P. 143-151.

41. *Thermodynamic properties of individual substances*, ed. V.P. Glushko et al. Vol. 3. 1981, Moscow: Nauka.

42. *De With, G.* High temperature fracture of boron carbide: experiments and simple theoretical models. // J. Mater. Sci. - 1984. - **19**, N 2. - P. 457−466.

43. *Nelmes, R.J., Loveday, J.S., Wilson, R.M., et al.* Observation of inverted-molecular compression in boron carbide. // Phys. Rev. Lett. - 1995. - **74**, N 12. - P. 2268.

Table 1. Normalized[‡] thermodynamic data of $B_6O$, other boron-rich solids and $\alpha$-$B_2O_3$

| Fitting parameters of Eq (1) for heat capacity | | | | | |
|---|---|---|---|---|---|
| Phases | $\theta_h$, K | $\theta_l$, K | $C_1$ | $C_2$ | $A$ |
| **$B_6O$** | **1020** | **1175** | **-0.0865** | **1** | **0.125** |
| $B_4C$ | | 1480 | | | |
| $\alpha$-$B_{12}$* | ~700 | 1300 | | | |
| $\beta$-$B_{106}$* | 1250 | 1700 | | | |
| Standard values of thermodynamic functions at 298.15 K | | | | | |
| Phase | $C_p^\circ$, J mol$^{-1}$ K$^{-1}$ | $H^\circ - H_0^\circ$, kJ mol$^{-1}$ | $S^\circ - S_0^\circ$, J mol$^{-1}$ K$^{-1}$ | $\Phi^\circ$, J mol$^{-1}$ K$^{-1}$ | |
| **$B_6O$** | **10.24** | **1.147** | **5.621** | **1.774** | |
| $B_4C$[†] | 10.62 | 1.122 | 5.422 | 1.658 | |
| $\alpha$-$B_{12}$* | 10.23 | 1.073 | 5.142 | 1.546 | |
| $\beta$-$B_{106}$[†] | 11.09 | 1.222 | 5.900 | 1.801 | |
| $\alpha$-$B_2O_3$[†] | 12.55 | 1.860 | 10.794 | 4.555 | |

‡ Normalized values correspond to the mole of atoms, not to the mole of formula units. The molar values of thermodynamic functions can be obtained by multiplying the corresponding normalized values on the number of atoms in the formula unit (e.g. 7 for $B_6O$).

\* Data from Ref. 40

† Data from Ref. 41

Table 2. Mechanical properties of $B_6O$, $B_4C$ and $\beta$-$B_2O_3$

|  | $B_6O$ | $\beta$-$B_2O_3$ | $B_4C$ |
|---|---|---|---|
| Hardness, $H_V$ (GPa) | 38 [5]; 40-45[§] [6] | 16 [32] | 45 [37] |
| Fracture toughness, $K_{Ic}$ (MPa·m$^{1/2}$) | 4.5* [6] |  | 3-4 [42] |
| Bulk modulus, $B_0$ (GPa) | 181 [26] | 170 [26] | 199 [43] |
| $B'_0$ | 6 [26] | 2.5 [26] | 1 [43] |
| Density, $\rho$ (g·cm$^{-3}$) | 2.620 | 3.111 | 2.516 |

[§] Overestimated values obtained at low indentation load (1 N).

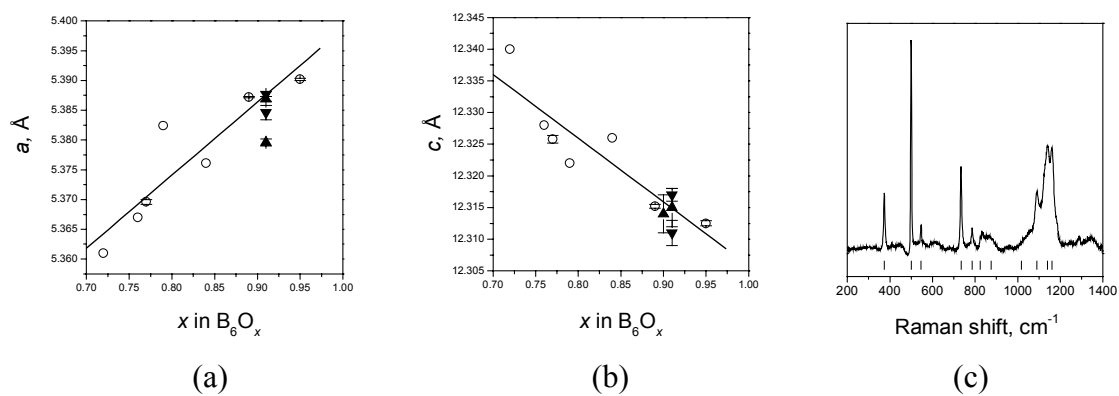

(a)                  (b)                  (c)

Fig. 1. Structural and lattice-vibration properties of $B_6O$: (**a** & **b**) lattice parameters of $B_6O_x$ as function of composition $x$ (symbols represent experimental data: ▲ and ▼ – our data, ○ – data from [16], lines are guides to the eye) and (**c**) Raman spectrum obtained using the 785-nm excitation light [18].

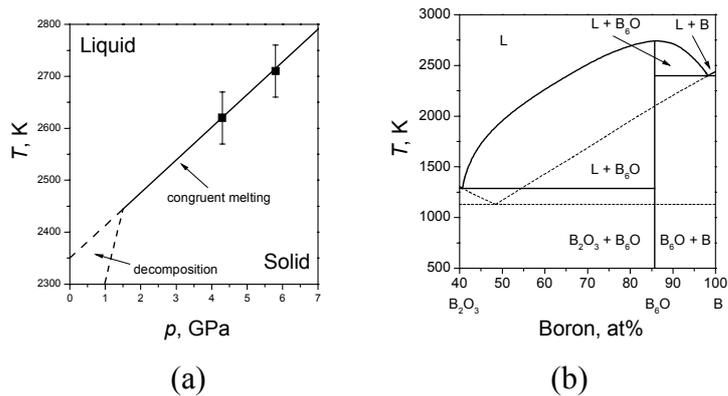

Fig. 2. Melting and thermal stability of $B_6O$: (**a**) melting curve of $B_6O$ (solid line; ■ - experimental data taken from Ref. 21) with decomposition region (dashed lines) [5] and (**b**) equilibrium (solid lines) and metastable (dotted lines) phase diagrams of the $B$–$B_2O_3$ system at 5 GPa [22].

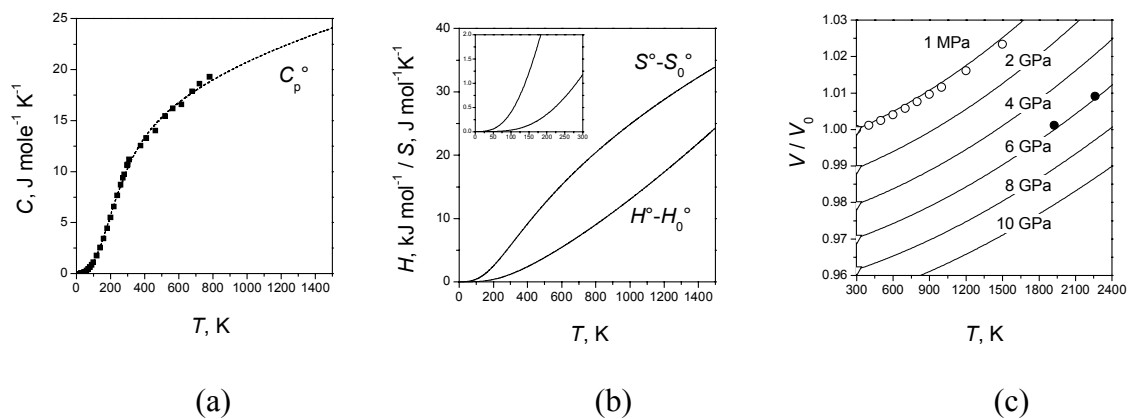

Fig. 3. Normalized values of thermodynamic functions of $B_6O$ (see note to Tab. 1): (**a**) heat capacity as a function of temperature (■ - experimental data taken from Ref. 23, dashed line - the fit to equation (1)), (**b**) enthalpy and entropy as functions of temperature (calculations were performed using fitted $C_p$ values), and (**c**) the relative volume-temperature isobars (symbols represent experimental data: ● – our data, ○ – data from Ref. 30, and ∇ – data from Ref. 26; while continuous lines show fit/prediction using equation (5)).

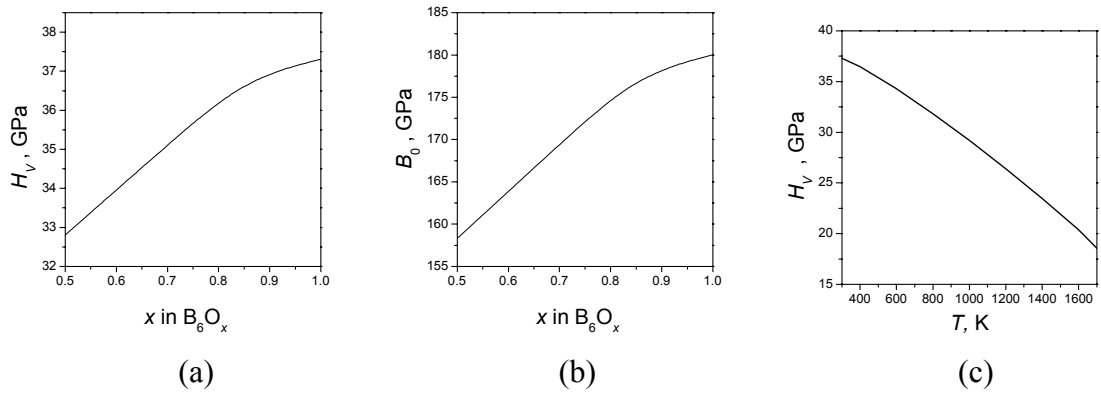

Fig. 4. Impact of composition *x* on Vickers hardness $H_V$ **(a)** and bulk modulus $B_0$ **(b)** of $B_6O_x$. Hardness of $B_6O$ as a function of temperature **(c)**. All lines represent simulations performed using equations (7) and (8).